\documentclass[showpacs,prb,twocolumn]{revtex4}
\usepackage{amsmath}
\usepackage{graphicx}
\usepackage{dcolumn}
\usepackage{bm}

\begin{document}
\title{Ground-state factorization and quantum phase transition in dimerized spin chains}

\author{Gian Luca Giorgi}\thanks{Present address: Institute for Cross-Disciplinary Physics and Complex Systems, IFISC (CSIC-UIB),
Campus Universitat Illes Balears, E-07122 Palma de Mallorca, Spain}\email{gianluca@ifisc.uib-csic.es}
\affiliation{Dipartimento di Fisica, Universit\`{a} di
Roma La Sapienza, Piazzale A. Moro 2, 00185 Roma, Italy}

\pacs{03.67.Mn, 75.10.Jm, 64.70.Tg}

\begin{abstract}
We study the occurrence of ground-state factorization in dimerized $XY$ spin
chains in a transverse field. 
Together with the usual ferromagnetic and
antiferromagnetic regimes, a third case emerges, with no analogous in
translationally-invariant systems, consisting of an antiferromagnetic
Ne\'{e}l-type ground state where pairs of spins represent the unitary cell. Then, we calculate the exact solution of the model and show that the factorizing field represent an accidental degeneracy point of the Hamiltonian. Finally, we extend the study of the existence of ground-state factorization to a more general class of models.
\end{abstract}
\maketitle

The study of zero-temperature critical phenomena in quantum magnetic systems
represents since long time a major research subject.\cite
{sachdev,takahashi,lieb,pfeuty} In particular, the $XY$ model is a very rich
source of information about the quantum behavior of spin chains because of
the availability of an exact analytical solution.\cite{lieb,barouch} During
last years, the main interest about spin chains concerned the relationship
between quantum phase transitions and entanglement \cite
{osborne,fazio,vidal,vedral}. Among a number of interesting properties of
such systems, it is worth citing the existence of special values of the
external magnetic field, the parameter which drives the phase transition,
which give rise to ground state factorization, discovered by Kurmann {\it et
al.}~.\cite{kurmann} This phenomenon has been observed in two-dimensional
lattices through quantum Monte Carlo methods \cite{verrucchi} and fully
analyzed by Giampaolo {\it et al.}\cite{giampaolo} in a recent publication~,
where the factorizing field has been determined for a quite general class of
models. They developed an appropriate measure of entanglement which vanishes
at the factorizing point. So far, the existence of a factorized ground state
has been predicted only in translationally invariant Hamiltonian models.

Moreover, critical properties of physical systems are discussed by taking
the thermodynamic limit from the beginning. On the other hand, the knowledge
of a finite-size solution clarifies important aspects of this limit. For
example, it is known that the quantum phase diagram of the $XY$ chain in a
transverse field exhibits two different symmetry-broken regions
characterized by different behaviors of two-body correlation functions. This dissimilarity has a microscopic origin easily
understood in the finite-size case. Besides these considerations, the study
of finite systems is relevant by itself for the realization of
mesoscopic qubits of contemporary interest.

In this Rapid Communication we discuss a finite-size dimerized $XY$ spin chain in a
transverse field, and analyze ground-state properties. The interest about
such system, belonging to a more general class of models,\cite
{tong,lima,derzhko} is motivated by experimental work on quasicrystals and
quasiperiodic superlattices.\cite{Shechtman,merlin} First of all, we show
that the model admits the existence of a factorized ground state, and then
discuss the exact solution. The factorizing field turns out to be an
accidental degeneracy point of the Hamiltonian and falls on a border surface
between two regions that, in the thermodynamic limit, are characterized by
different symmetry-breaking mechanisms. Furthermore, we will be able to
detect the conditions for the existence of ground-state factorization also
in a more general class of dimerized chains which are the generalization of
the model discussed in Ref. \cite{giampaolo}.

We start our discussion by considering a nearest-neighbor dimerized chain of
an even number $N$ of spin 1/2:
\begin{eqnarray}
{\cal H}&=&\sum_{l=1}^{N/2}\sum_{i=1}^{2}( \frac{J_{i}+\gamma _{i}}{2}%
\sigma _{2l-2+i}^{x}\sigma _{2l-1+i}^{x}\nonumber\\&+&\frac{J_{i}-\gamma _{i}}{2}\sigma
_{2l-2+i}^{y}\sigma _{2l-1+i}^{y}) -h\sum_{l=1}^{N}\sigma _{l}^{z},
\label{hamilton}
\end{eqnarray}
with  $\sigma _{N+1}^{\alpha }=\sigma _{1}^{\alpha }$, being $\sigma _{l}^{\alpha }$ the $\alpha $th Pauli matrix ($\alpha
=x,y,z $). Without loss of generality, we will limit the analysis to
positive fields. The Hamiltonian of Eq. (\ref{hamilton}) can be recast in the form of
a sum over two-body Hamiltonians ${\cal H}=\sum_{l=1}^{N/2}({\cal H}%
_{l}^{\left( 1\right) }+{\cal H}_{l}^{\left( 2\right) })$, where
\begin{eqnarray}
{\cal H}_{l}^{\left( i\right) }&=&\frac{J_{i}+\gamma _{i}}{2}\sigma
_{2l-2+i}^{x}\sigma _{2l-1+i}^{x}+\frac{J_{i}-\gamma _{i}}{2}\sigma
_{2l-2+i}^{y}\sigma _{2l-1+i}^{y}\nonumber\\&-&h_{i}\left( \sigma _{2l-2+i}^{z}+\sigma
_{2l-1+i}^{z}\right) ,
\end{eqnarray}
with $h_{1}$ and $h_{2}$ such that $h=h_{1}+h_{2}$.

The central feature of ${\cal H}_{l}^{\left( i\right) }$ is the invariance
under rotations of $\pi $ around the $z$ axis. This is formalized by the
vanishing of the commutator $[{\cal H}_{l}^{\left( i\right) },{\cal P}%
_{l}^{\left( i\right) }]=0,$ where ${\cal P}_{l}^{\left( i\right) }=\sigma
_{2l-2+i}^{z}\sigma _{2l-1+i}^{z}$ \ is the parity operator, since its
eigenvalues are $+1$ or $-1$, according to the number of down spins in the $%
z $ direction being even or odd. The above commutation relation then
requires also the eigenstates of ${\cal H}_{l}^{\left( i\right) }$ to have
definite parity. The problem is to establish whether there exists a set of
the Hamiltonian parameters such that the ground state is of the form $|\Psi
\rangle =\otimes _{l}|\psi _{l}\rangle .$ Notice that, if this is the case,
as remarked in Ref.~\cite{kurmann}, $|\psi _{2l-2+i}\rangle |\psi
_{2l-1+i}\rangle $ must be the ground state of ${\cal H}_{l}^{\left(
i\right) }$. The problem is then reduced to find the conditions under which $%
{\cal H}_{l}^{\left( i\right) }$ admits a factorized ground state. Now,
since ${\cal H}_{l}^{\left( i\right) }$ is not diagonal in the $\sigma ^{z}$
basis, if $|\psi _{2l-2+i}\rangle |\psi _{2l-1+i}\rangle $ has to be the
ground state, each factor must be of the form $\left| \psi _{l}\right\rangle
=\cos \psi _{l}\left| \uparrow \right\rangle +\sin \psi _{l}\left|
\downarrow \right\rangle ,$ with $\psi _{l}\neq 0,\pi /2$. This, in turn,
implies that in the factorized ground state the parity symmetry is broken
and, therefore, looking for the condition on the parameters of the
Hamiltonian leading to factorization of the ground state amounts to looking
for the condition leading to the degeneracy of the even and odd lowest lying
eigenstates, without invoking the vanishing of entanglement indicators.\cite
{giampaolo}

There are three different physical scenarios to be considered: (i) both $%
J_{1} $ and $J_{2}$ are negative (ferromagnetic case); (ii) both $J_{1}$ and $%
J_{2}$ are positive (antiferromagnetic case); (iii) $J_{1}>0$ and $J_{2}<0$
or vice versa (hybrid case). In any of these cases, factorization appears if
and only if $\gamma _{1}/J_{1}=\gamma _{2}$ $/J_{2}=\kappa $, i.e., only in
the presence of perfect dimerization of the longitudinal part of ${\cal H}$,
the factorized point falls in $h_{F}=\left[ \left( J_{1}+J_{2}\right) /2%
\right] \sqrt{1-\gamma ^{2}}$, and $\tan \psi _{2l-1+i}=\pm \left[ \left( 1-%
\sqrt{1-\kappa ^{2}}\right) /\kappa \right] ^{1/2}$.

(i) In the ferromagnetic case, we find $\psi _{2l-1+i}=\psi _{2l-2+i}$. As
expected, the factorized state is fully aligned along two possible
directions:\thinspace $\left| \Psi \right\rangle =\otimes _{l}\left| \psi
_{l}^{\pm }\right\rangle $.

(ii) In the second (antiferromagnetic) case, we find $\psi _{2l-1+i}=-\psi
_{2l-2+i}$. Then, alternate directions for the spin determining a
Ne\'{e}l-type ground state are observed: $\left| \Psi \right\rangle =\otimes
_{l=0}^{\left( N/2\right) -1}\left| \psi _{2l+1}^{\pm }\right\rangle \left|
\psi _{2l+2}^{\mp }\right\rangle $.

(iii) If both ferromagnetic and antiferromagnetic factorized ground states
are of the same kind of those obtained in the homogeneous $XY$ chain,\cite
{kurmann} the third (hybrid) case shows up an original character. Indeed,
by assuming, for example, $J_{1}<0$ and $J_{2}>0$ we find the constrains $%
\psi _{2l-1}=\psi _{2l}$ and $\psi _{2l}=-\psi _{2l+1}$. As a consequence,
the factorized ground state assumes the structure $\left| \Psi
\right\rangle =\otimes _{l=0}^{\left( N/4\right) -1}\left| \psi _{4l+1}^{\pm
}\right\rangle \left| \psi _{4l+2}^{\pm }\right\rangle \left| \psi
_{4l+3}^{\mp }\right\rangle \left| \psi _{4l+4}^{\mp }\right\rangle .$ Thus,
we obtain an antiferromagnetic Ne\'{e}l-type ground state, whose unitary
cell is represented by a pair of spins. An additional requirement for the
existence of the FP in this case is that $N/4$ must be an integer number in
order to avoid frustration effects.

Now, we discuss the general exact solution of Hamiltonian (\ref{hamilton}%
) and enlighten the role played by the factorizing field. The diagonalization
method is given in Ref. \cite{perk}. We discuss explicitly the finite-size
limit.\cite{lima,katsura} The first step is the introduction of the
Jordan-Wigner transformation, mapping spins into spinless fermions,\cite{lieb}
defined through $\sigma _{l}^{z}=1-2c_{l}^{\dagger }c_{l}$, $\sigma
_{l}^{+}=\prod_{j<l}\left( 1-2c_{l}^{\dagger }c_{l}\right) c_{l}$, and $%
\sigma _{l}^{-}=\prod_{j<l}\left( 1-2c_{l}^{\dagger }c_{l}\right)
c_{l}^{\dagger }$, which leads to ${\cal H}={\cal H}_{0}-{\cal PH}_{1}$,
with
\begin{eqnarray}
{\cal H}_{0} &=&\sum_{l=1}^{N/2}\sum_{i=1}^{2}[ J_{i}\left(
c_{2l-2+i}^{\dagger }c_{2l-1+i}+h.c.\right) \nonumber\\&+&\gamma _{i}\left(
c_{2l-2+i}^{\dagger }c_{2l-1+i}^{\dagger }+h.c.\right) ]
-h\sum_{l=1}^{N}\left( 1-2c_{l}^{\dagger }c_{l}\right) ,\nonumber\\ \\
{\cal H}_{1} &=&\left[ J_{2}\left( c_{N}^{\dagger }c_{1}-c_{N}c_{1}^{\dagger
}\right) +\gamma _{2}\left( c_{N}^{\dagger }c_{1}^{\dagger
}-c_{N}c_{1}\right) \right] ,
\end{eqnarray}
where the parity operator is ${\cal P}=\prod_{l=1}^{N}\left(
1-2c_{l}^{\dagger }c_{l}\right) $. Since $\left[ {\cal H},{\cal P}\right] =0$%
, all eigenstates of ${\cal H}$\ have definite parity, and we can proceed to
a separate diagonalization of ${\cal H}$ in the two subspaces corresponding
to ${\cal P}=\pm 1$. Then, the complete set of eigenvectors of ${\cal H}$
will be given by the odd eigenstates of\ ${\cal H}^{-}=$\ ${\cal H}_{0}+%
{\cal H}_{1}$ and the even eigenstates of ${\cal H}^{+}=$\ ${\cal H}%
_{0}-{\cal H}_{1}$. Both for ${\cal H}^{+}$\ and ${\cal H}^{-}$
the diagonalization can be performed through the division of the lattice in
two sublattices: $c_{2l-1}=a_{l}$, and $c_{2l}=b_{l}$, and with the help of
two separate Fourier transforms $a_{l}=\left( N/2\right)
^{-1/2}\sum_{k}a_{k}\exp [-i\frac{4\pi kl}{N}]$ and $b_{l}=\left( N/2\right)
^{-1/2}\sum_{k}b_{k}\exp [-i\frac{4\pi kl}{N}]$, where $k=0,1,\ldots \left(
N/2\right) -1$ in ${\cal H}^{-}$, and $k=1/2,3/2,\ldots \left( N/2\right)
-1/2$ in ${\cal H}^{+}$, getting
\begin{eqnarray}
{\cal H}^{\pm }&=&\sum_{k}\left[ J_{k}a_{k}^{\dagger }b_{k}-J_{k}^{\ast
}a_{k}b_{k}^{\dagger }+\gamma _{k}a_{k}^{\dagger }b_{-k}^{\dagger }-\gamma
_{k}^{\ast }a_{k}b_{-k}\right] \nonumber\\&-&h\sum_{k}\left( 2-2a_{k}^{\dagger
}a_{k}-2b_{k}^{\dagger }b_{k}\right) ,
\end{eqnarray}
with\ $J_{k}=J_{1}+J_{2}\exp [-i\frac{4\pi kl}{N}]$, $\gamma _{k}=\gamma
_{1}-\gamma _{2}\exp [-i\frac{4\pi kl}{N}]$. We remark that the difference
between ${\cal H}^{+}$ and ${\cal H}^{-}$ consists in the different set of
allowed values of $k$. Finally, we introduce a generalized Bogoliubov
transformation connecting $a_{k},a_{-k}^{\dagger },b_{k},b_{-k}^{\dagger }$,
and obtain two different kinds of quasiparticles,
\begin{equation}
{\cal H}^{\pm }=\sum_{k}\Lambda _{k}^{\left( +\right) }\left( 2\eta
_{k}^{\dagger }\eta _{k}-1\right) +\sum_{k}\Lambda _{k}^{\left( -\right)
}\left( 2\xi _{k}^{\dagger }\xi _{k}-1\right) ,
\end{equation}
where the eigenvalues, belonging to two branches separated by an energy gap,
are, for $k\neq 0,N/4$, $\Lambda _{k}^{\left( \pm \right) }=\sqrt{r_{k}\pm
\sqrt{s_{k}}}$, with $r_{k}=4h^{2}+\left| J_{k}\right| ^{2}+\left| \gamma
_{k}\right| ^{2}$ and $s_{k}=\left| J_{k}\right| ^{2}\left( 16h^{2}+2\left|
\gamma _{k}\right| ^{2}\right) +J_{k}^{2}\gamma _{k}^{\ast 2}+\gamma
_{k}^{2}J_{k}^{\ast 2}$, and $\Lambda _{0}^{\left( \pm \right) }=\sqrt{%
4h^{2}+\left( \gamma _{1}-\gamma _{2}\right) ^{2}}\pm \left(
J_{1}+J_{2}\right) ,$ $\Lambda _{N/4}^{\left( \pm \right) }=\sqrt{%
4h^{2}+\left( \gamma _{1}+\gamma _{2}\right) ^{2}}\pm \left(
J_{1}-J_{2}\right) $. Let us assume, for example, $\left( J_{1}+J_{2}\right) >0$
and $J_{1}>J_{2}$. Then, each $\Lambda _{k}$ is positive, with the exception
of $\Lambda _{0}^{\left( -\right) },\Lambda _{N/4}^{\left( -\right) }$ (both
of them are eigenvalues of ${\cal H}^{-}$), which can assume negative values,
respectively, for $h<h_{C}^{\left( 1\right) }=\sqrt{\left( J_{1}-J_{2}\right)
^{2}-\left( \gamma _{1}+\gamma _{2}\right) ^{2}}/2$ and for $h<h_{C}^{\left(
2\right) }=\sqrt{\left( J_{1}+J_{2}\right) ^{2}-\left( \gamma _{1}-\gamma
_{2}\right) ^{2}}/2$.\ Let us assume also, without loss of generality, $%
h_{C}^{\left( 1\right) }<h_{C}^{\left( 2\right) }$. Due to the foregoing
considerations, the ground state of ${\cal H}^{+}$ is its vacuum, and the
corresponding eigenvalue is $E_{0}^{+}=-\sum_{k=1/2}^{(N/2)-(1/2)}\sum_{\nu
=+}^{-}\Lambda _{k}^{\left( \nu \right) }$. As for ${\cal H}^{-}$, the
lowest energy is $E_{0}^{-}=-\sum_{k=0}^{(N/2)-1}\sum_{\nu =+}^{-}|\Lambda
_{k}^{\left( \nu \right) }|$, while the ground-state structure depends on $h$%
. The ground state has two quasiparticles on modes $k=0,N/4$\ for $%
h<h_{C}^{\left( 1\right) }$, one quasiparticle on the mode $k=0$\ for $%
h_{C}^{\left( 1\right) }<h<h_{C}^{\left( 2\right) }$, and finally, it is the
vacuum state for $h>h_{C}^{\left( 2\right) }$.

In order to identify the ground state of ${\cal H}$, we must compare the
lowest eigenvalue of ${\cal H}^{+}$ belonging to an even eigenstate ($%
E_{0}^{even}$) with the lowest eigenvalue of ${\cal H}^{-}$ belonging to an
odd eigenstate ($E_{0}^{odd}$). Since the vacuum state is even, we
immediately state that lowest even eigenvalue of ${\cal H}$ is $%
E_{0}^{even}= $ $E_{0}^{+}$. As far as the lowest odd eigenvalue of ${\cal H}
$ is considered, only inside the region $\{h_{C}^{\left( 1\right)
},h_{C}^{\left( 2\right) }\}$, where there is one quasiparticle (odd number
of excitations), $E_{0}^{odd}=E_{0}^{-}$. In fact, outside this region, $%
E_{0}^{-}$ belongs to even eigenstates (vacuum or two-quasiparticle state),
and we must look to the first excited state of ${\cal H}^{-}$. Thus, $%
E_{0}^{odd}(h>h_{C}^{\left( 2\right) })=E_{0}^{-}(h>h_{C}^{\left( 2\right)
})+\Lambda _{0}^{\left( -\right) }$, and $E_{0}^{odd}(h<h_{C}^{\left(
1\right) })=E_{0}^{-}(h<h_{C}^{\left( 1\right) })+\Lambda _{N/4}^{\left(
-\right) }$.

In the thermodynamic limit, the sum over $k$ becomes an integral, and the
vacuum energies of ${\cal H}^{+}$ and ${\cal H}^{-}$ are identical: $%
E_{0}^{+}=E_{0}^{-}$. Thus, for $h_{C}^{\left( 1\right) }<h<h_{C}^{\left(
2\right) }$, $E_{0}^{odd}=E_{0}^{even}$ and spontaneous symmetry breaking
comes out. Outside this range, the ground state has definite parity, being
the energy gap $\Delta E=2\Lambda _{0}^{\left( -\right) }$ for $h\geq
h_{C}^{\left( 2\right) }$ and $\Delta E=2\Lambda _{N/4}^{\left( -\right) }$
for ${\rm \ }h\leq h_{C}^{\left( 1\right) }$. Thus, although phase
transitions take place only for macroscopic systems, the change in the
energy sign of quasiparticles can be used as precursive property, allowing
one to characterize the transition before performing the thermodynamic limit.

\begin{figure}
  \includegraphics[height=9cm,angle=90]{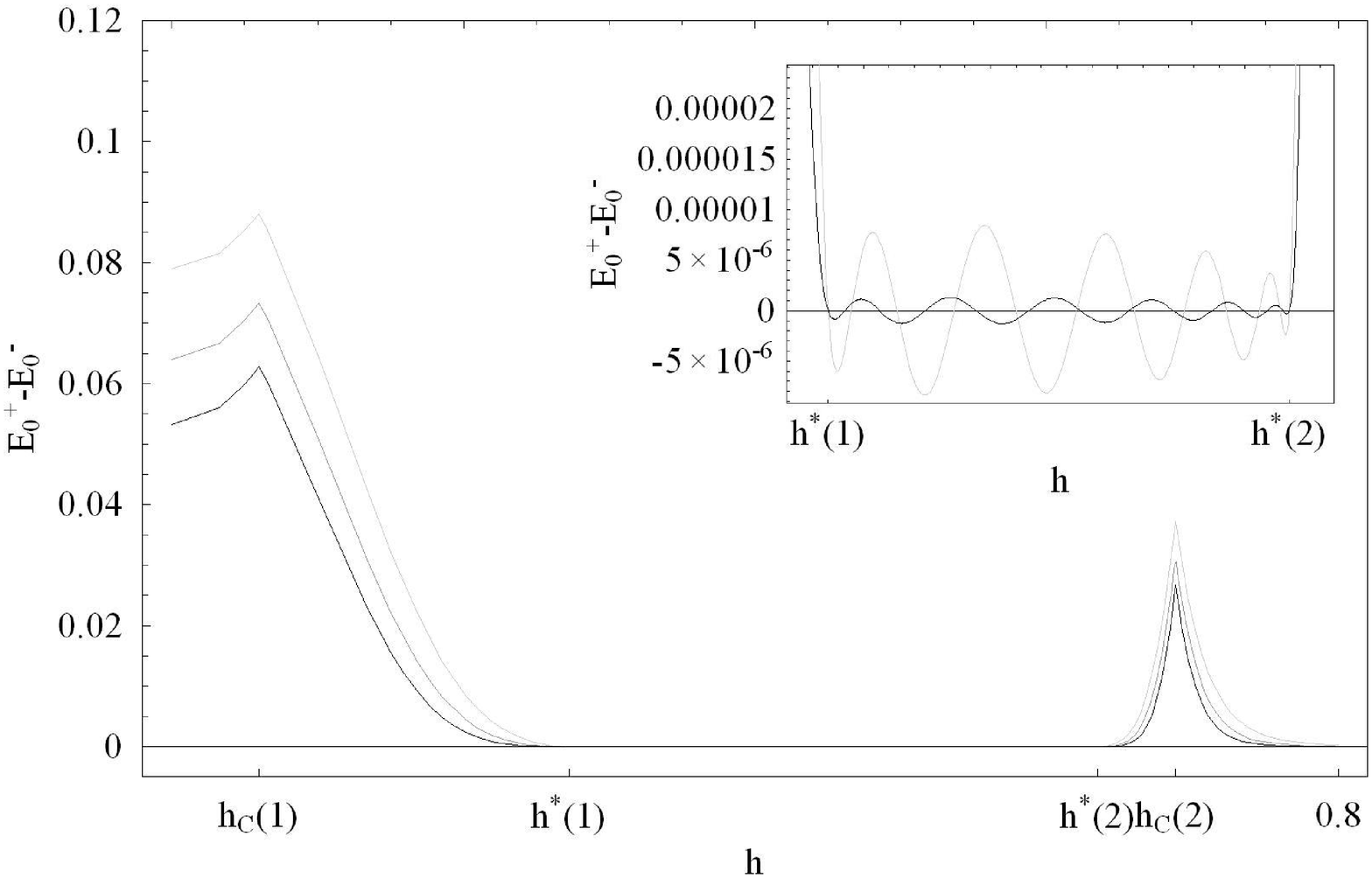}\\
  \caption{The energy difference $E_{0}^{+}-E_{0}^{-}$ between the two
ground states of ${\cal H}^{+}$ and ${\cal H}^{-}$. Inside the range $%
\{ h_{C}^{\left( 1\right) },h_{C}^{\left( 2\right) }\} $\ these
two states are the lowest eigenstates of ${\cal H}$.
The dimensionless Hamiltonian parameters are $J_{1}=1,J_{2}=0.4,\gamma
_{1}=0.42,\gamma _{2}=0.168$. These choice is compatible with the existence
of the factorized ground state ($h_{F}=h^{\ast }\left( 2\right) $), which is
found to be around $h=0.63$. The other relevant parameters are $%
h_{C}^{\left( 1\right) }\simeq 0.06$,  $h^{\ast }\left( 1\right) \simeq 0.27$%
, and $h_{C}^{\left( 2\right) }\simeq 0.69$. The light gray line refers to a chain of 20 spins, the gray line corresponds to 24 spin, while for the black curve a chain of 28 spin has been used. In the inset we plot a detail of the bigger picture in the case $N=24$ (light gray) and $N=28$ (black) to enlighten the level crossing. As expected, we
found $N/2$ intersection points for each curve.}\label{figura}
\end{figure}

As in the homogeneous chain,\cite{barouch} in the presence of spontaneous
symmetry breaking, two-body correlation functions can decrease monotonically
or oscillate as a function of the spin distance depending on the Hamiltonian
parameters. Recently, de Lima {\it et al.} \cite{lima} found different
regions in the phase diagram separated by hypersurfaces (collapsing into
lines when $\gamma _{1}/J_{1}=\gamma _{2}$ $/J_{2}$). We show that the first
one of the separating lines corresponds to the factorizing field. Indeed,
the different behavior of the correlation functions derives from different
symmetry-breaking mechanisms. By analyzing lowest odd and even eigenvalues
of ${\cal H}$ in the symmetry broken region for any finite $N$ (Fig. \ref
{figura}), we observe a series of $N/2$\ intersection point $h_{i}$. The
existence of such points has been discussed for the homogeneous chain in
Refs. \cite{hoeger} ,and \cite{rossignoli} and is responsible for magnetization jumps of Ref.
\cite{puga}. If $\gamma _{1}/J_{1}=\gamma _{2}/J_{2}$, the first crossing
point (for decreasing fields) is at $h=h_{F}$ for any $N$. In fact, $%
\sum_{k=0}^{N/2-1}\sum_{\nu =+}^{-}\Lambda _{k}^{\left( \nu \right) }\left(
h_{F}\right) =\sum_{k=1/2}^{N/2-1/2}\sum_{\nu =+}^{-}\Lambda _{k}^{\left(
\nu \right) }\left( h_{F}\right) =N(J_{1}+J_{2})$. That is, $h=h_{F}$, is an
accidental degeneracy point. If $\gamma _{1}/J_{1}\neq \gamma _{2}$ $/J_{2}$%
, there are not fixed points. However, with the increasing of $N$, all the
crossing point are confined inside the region limited by $h^{\ast }\left(
1\right) =\sqrt{\left( J_{1}-J_{2}\right) ^{2}-\left( \gamma _{1}-\gamma
_{2}\right) ^{2}}/2$ and $h^{\ast }\left( 2\right) =\sqrt{\left(
J_{1}+J_{2}\right) ^{2}-\left( \gamma _{1}+\gamma _{2}\right) ^{2}}/2$.
These critical values define the separating surfaces of Ref. \cite{lima}. It
is worth noting that when a factorized field exists, $h_{F}=h^{\ast }\left(
2\right) $.

In the thermodynamic limit, this kind of structure implies two different
symmetry breaking mechanisms. As one can deduce from the results of Fig. \ref
{figura}, for $h^{\ast }\left( 1\right) <h<h^{\ast
}\left( 2\right) $, as $N\rightarrow \infty $, the set $\left\{
h_{i}\right\} $ of the degeneracy points becomes a denumerable infinity, i.e., an infinite number of crossing points appears and the two lower
 energies coincide, while for $h_{C}^{\left( 1\right) }<h<h^{\ast }\left( 1\right) $ and for $%
h^{\ast }\left( 2\right) <h<h_{C}^{\left( 2\right) }$ there is the usual symmetry
breaking due to the vanishing of the gap, i.e. the ground state has a definite parity for any finite $N$, but this difference goes to zero with $1/N$.   This is the microscopic
mechanism responsible for dissimilar two-body correlation functions. Then,
like in the homogeneous case,\cite{barouch} as the factorizing field is
reached, correlation functions change character.

To conclude, we extend the study of ground-state factorization to a larger
class of $XYZ$ dimerized long-range spin models: ${\cal H}=\sum_{r}{\cal H}_{r}$, with
\begin{eqnarray}
\mathcal{H}_{r}&=&\sum_{\alpha =x,y,z}\left[ \sum_{l=1}^{\left( N/2\right)
-1}J_{r,1}^{\alpha }\sigma _{2l-1}^{\alpha }\sigma _{2l-1+r}^{\alpha
}+\sum_{l=1}^{N/2}J_{r,2}^{\alpha }\sigma _{2l}^{\alpha }\sigma
_{2l+r}^{\alpha }\right]\nonumber\\ &-&h\sum_{l=1}^{N}\sigma _{l}^{z},  \label{longrange}
\end{eqnarray}
where $J_{r,i}^{\alpha }$ are the dimerized coupling constants between spin pairs
at odd distance $r$. The existence of alternate coupling on even
distances cannot be univocally introduced. However, we could consider
homogeneous coupling for such distances. Notice that, assuming $%
J_{r,1}^{\alpha }=J_{r,2}^{\alpha }$ for any $r$,\ we recover the class of
models considered by Giampaolo {\it et al.} in Ref.~ \cite{giampaolo}. For
the sake of clarity, we will consider explicitly the case of a
one-dimensional lattice, but the generalization to higher dimensions is
straightforward.

Let us first note that, to circumvent frustration effects, $\left(
N/r\right) $ has to be an integer number for any value of $r$ appearing in $%
{\cal H}$. As already done for the short-range $XY$ chain, we rewrite ${\cal H}%
_{r}$ as a sum of two-body Hamiltonians, with
\begin{eqnarray}
{\cal H}_{l,r}^{\left( i\right) } &=&J_{r,i}^{x}\sigma _{2l-2+i}^{x}\sigma
_{2l-2+i+r}^{x}+J_{r,i}^{y}\sigma _{2l-2+i}^{y}\sigma _{2l-2+i+r}^{y}
\nonumber \\
&+&J_{r,i}^{z}\sigma _{2l-2+i}^{z}\sigma _{2l-2+i+r}^{z}-h_{i,r}\left(
\sigma _{2l-2+i}^{z}+\sigma _{2l-2+i+r}^{z}\right) ,\nonumber\\
\end{eqnarray}
and $h_{1,r}$, $h_{2,r}$ such that $h=\sum_{r}\left( h_{1,r}+h_{2,r}\right) $.

The simplest case to study is the full ferromagnetic picture, where all
coupling constants are negative. By following the same procedure introduced
above, we proceed to calculate the ground-state energies for the two parity
subspaces in each of two kinds of dimers ($\mathcal{H}_{l,r}^{\left(
1\right) }$ and $\mathcal{H}_{l,r}^{\left( 2\right) }$), and force symmetry
breaking. The factorized ground state of each of such dimers amounts to
be $\left| \Psi \right\rangle _{l,r}^{\left( i\right) }=\left( \cos \psi
_{r}^{\left( i\right) }\left| \uparrow \right\rangle \pm \sin \psi
_{r}^{\left( i\right) }\left| \downarrow \right\rangle \right) _{l}\otimes
\left( \cos \psi _{r}^{\left( i\right) }\left| \uparrow \right\rangle \pm
\sin \psi _{r}^{\left( i\right) }\left| \downarrow \right\rangle \right)
_{l+r}$, with
\begin{equation}
\tan ^{2}\psi _{r}^{\left( i\right) }=\frac{%
J_{r,i}^{x}+J_{r,i}^{y}-2J_{r,i}^{z}+2\sqrt{\left(
J_{r,i}^{z}-J_{r,i}^{x}\right) \left( J_{r,i}^{z}-J_{r,i}^{y}\right) }}{%
J_{r,i}^{x}-J_{r,i}^{y}}
\end{equation},
The existence of a globally factorized ground state implies that the angle $%
\psi _{r}^{\left( i\right) }$ must be the same for any of the dimers
involving each site, i.e. it has to be independent both on $r$ and $i$. This
result is achieved by the following conditions: $J_{r,2}^{\alpha }=\kappa
J_{r,1}^{\alpha }$, and $J_{r,i}^{\alpha }=\gamma _{r}J_{1,i}^{\alpha }$.
The value of the factorizing field is then $h_{F}=\left( 1+\kappa \right)
\sqrt{\left( \mathcal{J}_{1}^{z}-\mathcal{J}_{1}^{x}\right) \left( \mathcal{J%
}_{1}^{z}-\mathcal{J}_{1}^{y}\right) },$where the $\mathcal{J}^{\alpha }$
are the global interactions along different axes: $\mathcal{J}_{i}^{\alpha
}=\sum_{r}J_{r,i}^{\alpha }$.

For the full antiferromagnetic case, let first assume the nearest-neighbor
interaction parameters such that $\left( J_{1,i}^{x}+J_{1,i}^{y}\right) >0$.
Then, we expect for the factorized ground state the structure $\left| \Psi
\right\rangle =\otimes _{l=0}^{N/2-1}\left| \psi _{2l+1}^{\pm }\right\rangle
\left| \psi _{2l+2}^{\mp }\right\rangle $. The range-dependent coupling
constants have to be compatible with the existence of this state. This is
possible only if $\left( J_{r,i}^{x}+J_{r,i}^{y}\right) >0$ for any $r$. The conditions that ensure the existence of a factorized ground state
are $J_{r,2}^{\alpha }=\kappa J_{r,1}^{\alpha }$, and $J_{r,i}^{\alpha
}=\gamma _{r}J_{1,i}^{\alpha }$, and the factorizing field amounts to $%
h_{F}=\left( 1+\kappa \right) \sqrt{\left( {\cal J}_{1}^{z}+{\cal J}%
_{1}^{x}\right) \left( {\cal J}_{1}^{z}+{\cal J}_{1}^{y}\right) }$, with $%
{\cal J}_{1}^{\alpha }=\sum_{r}J_{r,1}^{\alpha }$.

The second, hybrid, way to introduce antiferromagnetism is to fix, for
example, $\left( J_{1,1}^{x}+J_{1,1}^{y}\right) >0$ and $\left(
J_{1,2}^{x}+J_{1,2}^{y}\right) <0$, obtaining $\left| \Psi \right\rangle
=\otimes _{l=0}^{\left( N/4\right) -1}\left| \psi _{4l+1}^{\pm
}\right\rangle \left| \psi _{4l+2}^{\pm }\right\rangle \left| \psi
_{4l+3}^{\mp }\right\rangle \left| \psi _{4l+4}^{\mp }\right\rangle $ as a
factorized ground state. This implies $J_{r,2}^{x,y}=-\kappa J_{r,1}^{x,y}$,
$J_{r,2}^{z}=\kappa J_{r,1}^{z}$, $J_{r,i}^{z}=\gamma _{r}J_{1,i}^{z}$, and $%
J_{r,i}^{x,y}=\left( -1\right) ^{\left( r-1\right) /2}\gamma
_{r}J_{1,i}^{x,y}$, leading for the factorized field to $h_{F}=\left(
1+\kappa \right) \sqrt{\left( {\cal J}_{1}^{z}+{\cal J}_{1}^{x}\right)
\left( {\cal J}_{1}^{z}+{\cal J}_{1}^{y}\right) }$, with ${\cal J}%
_{1}^{z}=\sum_{r}J_{r,1}^{z}$ and ${\cal J}_{1}^{x,y}=\sum_{r}\left(
-1\right) ^{\left( r-1\right) /2}J_{r,1}^{x,y}$.

In summary, we studied the zero-temperature phase diagram of the dimerized
$XY$ chain in a transverse field. We discussed the existence of a fully
unentangled ground state, which depends on whether the parameters of the
system satisfy given property. Furthermore, we showed the role of the
factorizing field inside the general solution of the model. It represents a
border line between two separate symmetry-broken regions in the space of the
Hamiltonian parameters. In analogy with the homogeneous case, where these
two regions are characterized by qualitatively different types of
entanglement, namely parallel and antiparallel entanglements,\cite{amico} we
expect that the same transition could take place also in our system. On the
other hand, also when the Hamiltonian parameters are not compatible with
ground-state factorization, there is a value of the field which separates
the two regions. Finally, we extended the search for ground state
factorization to more general dimerized models.

The author gratefully acknowledges F. de Pasquale and M. Zannetti for
invaluable support.

\end{document}